\documentclass[10pt,letterpaper]{article}
\usepackage{opex3,cite}
\usepackage{amsmath, amssymb,siunitx}
\begin{document}
\newcommand{\Grad}{\Gamma_\text{rad}}
\newcommand{\Gradone}{\Gamma_\text{rad}^1}
\newcommand{\Gradtwo}{\Gamma_\text{rad}^2}
\newcommand{\Gradi}{\Gamma_\text{rad}^i}
\newcommand{\Grado}{\Gamma_\text{rad}^o}
\newcommand{\Gdiss}{\Gamma_\text{diss}}
\title{Critical coupling to Tamm plasmons}
\author{Baptiste Augui\'e$^*$, Axel
  Bruchhausen, Alejandro Fainstein}
\address{Centro At\'omico Bariloche e Instituto Balseiro, San Carlos de Bariloche, 8400 R\'io Negro, Argentina}
\email{$^*$baptiste.auguie@cab.cnea.gov.ar} 
\begin{abstract}
The conditions of critical coupling of light to Tamm plasmons are
investigated with comprehensive numerical simulations, highlighting
the parameters that maximise absorption of incident light in the
metal layer. The asymmetric response in reflection and absorption with
respect to the direction of incidence is discussed, the two cases
yielding different optimal coupling conditions. These findings are relevant for the design of optimised Tamm structures, particularly in
applications such as narrow-band thermal emitters, field-enhanced
spectroscopy and refractive-index sensing.
\end{abstract}
\bibliography{references}
\section{Introduction}
%
Tamm plasmons (TP) are electromagnetic modes confined between a
Distributed Bragg Reflector (DBR) and a noble metal (e.g. gold)~\cite{Gas03,Kal07,Sas10}; in
contrast to surface-plasmons they may be excited at normal incidence
and present a relatively high quality factor, triggering interest as a convenient platform for enhanced light-matter
interaction at the nanoscale~\cite{Sym09 ,Leo12,Sym12,Zha13,Xue13,Lee13,Afi13,Aug14}. Upon excitation of TPs with light, a dip in reflectivity
is observed, which may reach 0\% (critical coupling
condition) for optimised structures, whereby optical energy is distributed between transmission and
absorption channels. Many designs of plasmonic nanostructures have been proposed with the
purpose of harnessing visible and infra-red light with greater
efficiency~\cite{Cui14}. Potential applications include solar cells,
display and lighting technologies, optical sensors, and, by
virtue of Kirchhoff's law, thermal emitters~\cite{Lee05}. The planar
design of the Tamm structure, of great simplicity and amenable to
large-scale fabrication, offers a clear advantage over more complex,
3-dimensional plasmonic nanostructures. This work aims to present a simple and accurate physical
picture for optimising the coupling of incident light to TPs in realistic structures, supported by insights from the theory of critical coupling and extensive numerical
simulations.
\subsection{Critical coupling of light to open resonators}
A Tamm structure may be regarded as a dissymmetric cavity, enclosed by
a metal mirror on one side, and a DBR on the other. Based on the general theory of open resonators~\cite{Bli08}, resonant transmittance and
reflectance take the form,
\begin{equation}
T = \frac{4\Gradone\Gradtwo}{\left(\Gradone+\Gradtwo+\Gdiss\right)^2},\qquad
R = \frac{\left(\Gradtwo - \Gradone + \Gdiss\right)^2}{\left(\Gradone + \Gradtwo + \Gdiss\right)^2}
\label{eq:R}\end{equation}
where $\Gdiss$ is the damping factor due to intrinsic (ohmic) loss, $\Gradone$ and
$\Gradtwo$ the radiative loss factors for the two barriers. The transmittance $T$ may reach unity only for
non-absorbing, symmetric structures ($\Gdiss=0$, $\Gradone = 
\Gradtwo$). This situation occurs in dielectric cavities surrounded by
two identical DBRs; in this case the response of the cavity improves
monotonically with the reflectivity of the two
mirrors. The quality factor increases with the number of layers,
yielding a higher field enhancement inside the cavity, and maintaining complete
transmission at resonance. The introduction of a lossy metallic
element drastically changes the situation~\cite{Bli08}. Unit transmission is
no longer possible, as light undergoes absorption while interacting
with the lossy structure. Reflection may still reach 0\%, under the critical coupling condition $\Gdiss=\Gradone -\Gradtwo$. When 
transmission is prohibited ($\Gradtwo=0$), e.g. by way of total
internal reflection~\cite{Her94}, critical coupling coincides with complete light absorption. 
The precise balance between radiative and non-radiative loss
channels gives rise to a rich behaviour in the optical response, when
varying the structural parameters and incidence conditions. Previous
works have described complete absorption in similar planar
multilayers; in fact original studies preceding what is
now labelled a ``Tamm plasmon'' structure considered the related problem of maximising
thermal emissivity in a DBR-absorber structure~\cite{Neu09},
supporting surface phonon-polaritons. Perfect light absorbers based on a
DBR-lossy layer combination have also been proposed, using J-aggregates~\cite{Tis06},
nanoparticle composites~\cite{Vet13}, graphene~\cite{Pip14}, or metal films~\cite{Du10,Gon11}. However, to the best
of our knowledge, no systematic review of the critical coupling
mechanism with Tamm plasmons has been presented thus far. The present article aims to cover
this topic; in particular we describe the conditions that yield vanishing
reflectance for the Tamm structure, and further discriminate the regime of complete light absorption.
\subsection{Review of the Tamm plasmon resonance}
\begin{figure*}[htbp]
\centering
\includegraphics[width=\columnwidth]{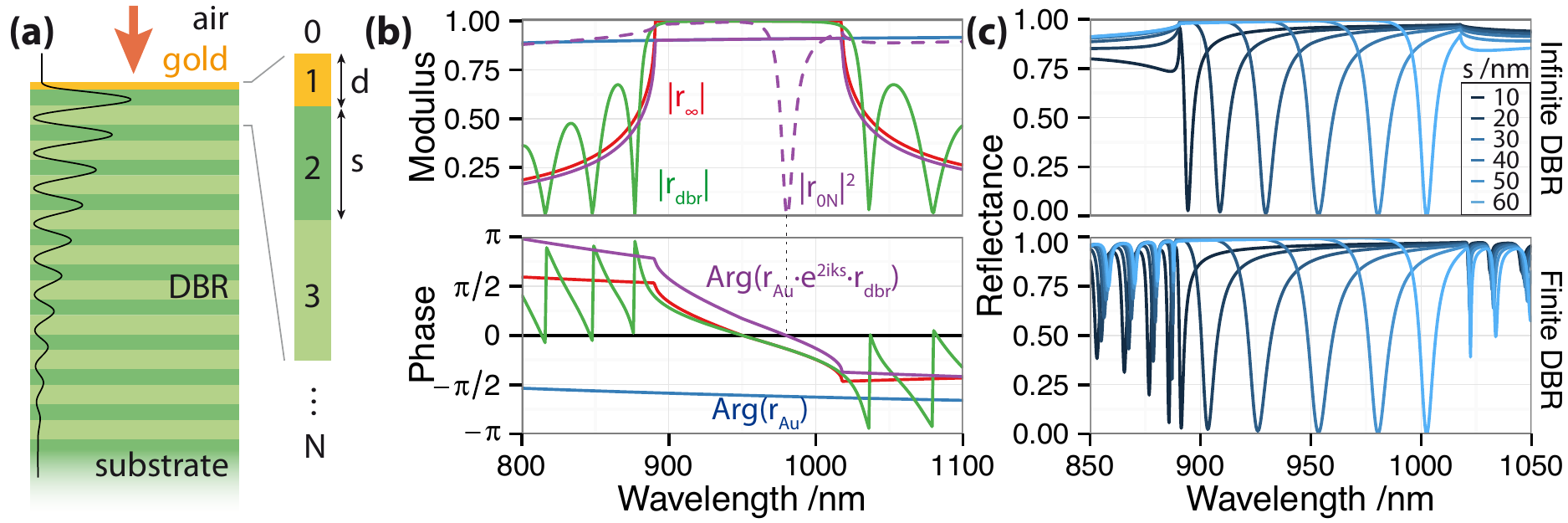}
\caption{(a) Schematic view of the Tamm structure; a GaAs/AlAs DBR is terminated
  by a thin gold film. (b) Components of the reflectivity complex
  variable for the Tamm structure, as described in the text. The top
  panel presents the modulus, the bottom panel the argument.  (c)
  Simulated reflectivity spectra for a Tamm structure with
  semi-infinite DBR (top), and finite DBR (bottom) with 50 pairs of
  layers, varying the thickness $s$ of the GaAs spacer adjacent to the Au metal layer from \SI{10}{nm} to \SI{60}{nm}.}
\label{fig0}
\end{figure*}
The Tamm structure is presented in Fig~\ref{fig0} (a). Although the
discussion would equally apply to other materials and spectral range
of operation, we will focus here on an experimentally-relevant
configuration in the near-infrared region, using AlAs and
GaAs for the DBR layers (refractive index $n_L=3.0$
and $n_H=3.7$, respectively, and quarter-wave layer thickness
for an operating wavelength of 950\,nm), and the complex dielectric function for
the gold layer obtained from Ref.~\cite{Ru09}. We note that in this
long-wavelength regime, away from the region of interband transitions,
gold closely follows the behaviour of a free-electron
metal and is also well described by a simple Drude model. To obtain the resonance condition associated with TPs, we take the pragmatic
approach of looking for a pole in the complex reflectivity coefficient
associated with the multilayer structure. We follow here a recursive
treatment based on the application of the Fresnel equations at
successive interfaces~\cite{Ami13}. The multilayer is indexed from 0
to N as shown in Fig~\ref{fig0} (a); the reflectivity of an interface
between layers (i,j) is labeled $r_{ij}$. Following a recursive approach,
the complex reflectivity may thus be decomposed as
\begin{equation}
		r_{\text{0N}}=\frac{r_{01}+r_{1N}\exp(2i
                  kn_{Au}d)}{1+r_{01}r_{1N}\exp(2i kn_{Au}d)}, \qquad 
		r_{\text{1N}}=\frac{r_{12}+r_{2N}\exp(2i
                  kn_{H}s)}{1+r_{12}r_{2N}\exp(2i kn_{H}s)},
\label{eq:r01}
\end{equation}
with $k=2\pi/\lambda$ the wavevector in a vacuum, $d$ the gold
thickness, $s$ the thickness of the GaAs spacer adjacent to
the metal. We seek a resonant term for the Au--DBR interface, which corresponds to
the second iteration, $r_{1N}$. The denominator of this expression yields the resonance condition for
TPs,
\begin{equation}
1-r_{\text{Au}}r_{\text{\sc dbr}}\exp(2i kn_{H}s) \approx 0,
\label{eq:pole}
\end{equation}
where we identified $r_{2N}=r_{\text{\sc dbr}}$ with the reflection coefficient of
the DBR (seen from the 2--3 interface), and $r_{\text{Au}}=-r_{12}$ with the
reflectivity of a Au--GaAs interface. This expression may also
be understood as the condition of constructive round-trip for light 
inside the cavity formed by the first GaAs layer sandwiched between
the metal and the rest of the DBR~\cite{Kal07}. In this sense, it characterises the
Tamm mode, independently of the condition of excitation, and therefore
also applies to the reverse configuration where light incident from a
finite DBR couples to the Tamm mode at the far end of the
structure. In this case, however, coupling conditions may
differ---$\Gamma_\text{rad}$ in particular is associated, in the
view of coupled-mode theory~\cite{Hau78}, with the overlap integral between modal and incident fields, the profile of the latter being
affected by the direction of incidence in the extended DBR.

Figure~\ref{fig0}(b) illustrates the contribution of the
different terms in Eq.~\ref{eq:pole}, and their relation to the TP
resonance. The top panel presents the modulus of the reflectivity
coefficient for a bare DBR, semi-infinite (red line), and finite (50 pairs
of layers, green line) with incidence from air. The characteristic stop band centred at 950\,nm is clearly
observed, and side-bands appear on both tails for the finite structure
as a result of Fabry-Perot resonances between multiple pairs of
layers~\cite{Mac10}. The reflectivity coefficient for a Au--GaAs interface is
shown in blue, with a relatively constant modulus over this spectral
range. The dashed purple curve presents for comparison the full
reflectance $R_\text{0N}=|r_\text{0N}|^2$ obtained from
Eq.~\ref{eq:r01}, the TP mode appearing as a sharp dip at 980\,nm. The bottom panel of Fig.~\ref{fig0}(b) presents the
complex argument of the same terms. The reflection coefficient for the
dielectric-Au interface produces a relatively constant phase shift of
about $-\pi/2$, while the DBR presents a linear phase shift across the stopband. The resonance condition expressed in
Eq.~\ref{eq:pole} requires
$\mathrm{Arg}\left(r_{\text{Au}}r_{\text{\sc dbr}}\exp(2i
  kn_{H}s)\right)= 0$; this zero-crossing point is observed around
980\,nm, and coincides with the reflectance minimum for
$|r_\text{0N}|^2$ in the top panel. 

\begin{figure*}[!Hbpt]
\centering
\includegraphics[width=\columnwidth]{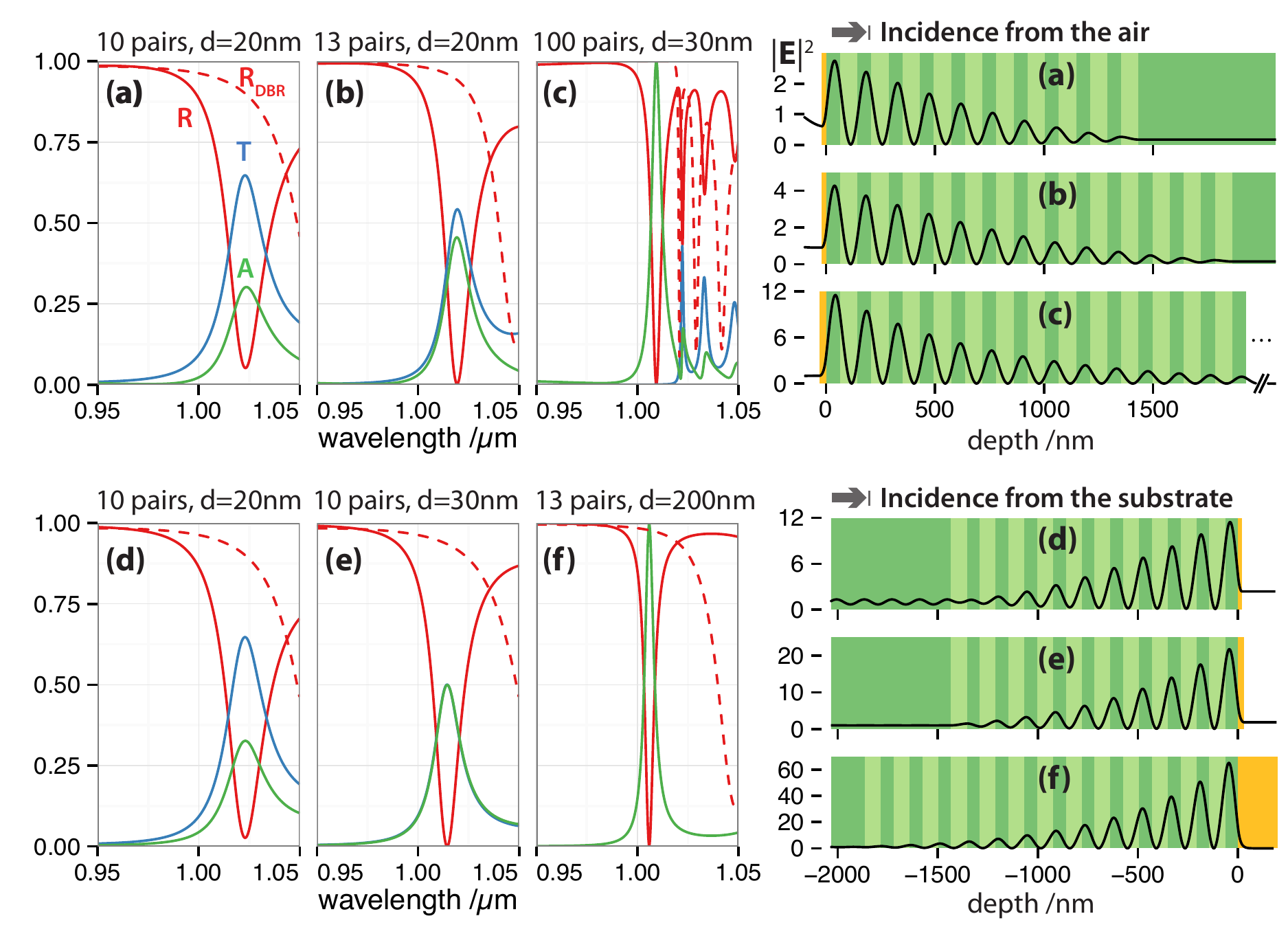}
\caption{Far-field response and mode profiles for different Tamm
  structures. Top panels consider light incident from the substrate--DBR side, while bottom panels consider the reverse situation of
  incidence from the air--gold side. (a, d) Unoptimised
  structure. (b,e) Critical coupling. (c,f) Full absorption. Note that
  the structure is displayed such that light always comes from the left side.}
\label{fig1}
\end{figure*}

A further interesting feature of TPs is the possibility of tuning the
resonance position across the stopband~\cite{Kal07}, by varying the spacer thickness
$s$, thereby affecting the phase-shift in Eq.~\ref{eq:pole}. This is further illustrated in Fig.~\ref{fig0}(c),
with numerical simulations of the reflectance for a semi-infinite Tamm
structure (top panel), and finite (50 pairs, bottom panel), where $s$
was varied from $10$\,nm to $60$\,nm. We note the
very close agreement in the TP resonance position and lineshape between
the two models, the only noticeable difference being the appearance of
side bands for the finite structure.

With these preliminary considerations in mind, we are now in a position to
discuss the coupling of light to the Tamm mode, which, as we shall
see, will crucially depend on the DBR parameters~\cite{Sas08,Du10}, the metal
thickness~\cite{Tsa11,Lee13, Zha14}
and dielectric function~\cite{Bra09a,Vet13}, as well as the direction of incidence~\cite{Bad14,Bad14a}. To simplify the
discussion, we will restrict the study to normal incidence, the
S.I. presents for completeness a simulation of the angular dispersion of the TP and its
effect on critical coupling, and we refer the reader to a recent study
for the relation between surface impedance and angular dispersion in
perfect absorbers~\cite{Tit14}.

Figure~\ref{fig1} provides a global overview of the various situations under
consideration.  The top panels (a,b,c) consider incidence from the air--Au
side, with the leftmost panels presenting the far-field optical
response (reflectance $R$, absorbance $A$, transmittance $T$) at normal incidence for three different Tamm structures, while the rightmost panels
show the mode profile at resonance (electric field intensity,
$|\mathbf{E}|^2$, normalised to the incident field). Three situations are considered,
(a) unoptimised structure, where light only partially couples to the
Tamm mode ($R\neq 0$); (b) \emph{critical coupling} situation with
$R=0$; (c) perfect absorption ($A=1$). The same layout is adopted for
the bottom panels (d, e, f), where structures were selected to
illustrate the reverse case of light incident from the substrate--DBR
side. This allows direct comparison between the two possible
experimental configurations. Note that the structure is mirrored in panels (d--f) so that
light is always incident from the left side.
A number of interesting features can be noted from these
simulations. First, a broad range of parameters (number of layers,
$N$, metal thickness $d$, spacer thickness $s$) can be tuned to reach a
critical coupling situation (b, c, e, f), with a different ratio of
absorbance and residual transmittance. Second, the reflectivity of
both mirrors (Au layer and DBR)  affects the spectral position of the
TP, as well as the linewidth of the mode. As the DBR (resp. Au) thickness
increases, the TP mode blue-shifts and gets narrower, as a consequence of
reduced radiative damping. Third, we focus on the influence of the
direction of incidence on the optical response. Cases (a) and (e) are
for the exact same structure, only the direction of incidence
differs. The far-field properties are qualitatively similar, with a
negligible shift of the TP resonance for these specific parameters. The
electric field in (a) and (d) appears more strongly enhanced when light is incident from the substrate side, but
this is simply a consequence of the incident irradiance (flux of
the Poynting vector) scaling with the refractive index of the incident
medium ($n_1=3.7$ for the GaAs substrate).

The asymmetry in
the response is better revealed in the different \emph{optimised}
conditions to reach critical coupling and perfect absorption. When
light is incident from the air-Au side, an optimum and finite metal
thickness is required to allow coupling of incident light with the
Tamm plasmon mode. The DBR, however, may extend to infinity, as it
prevents radiative leakage through transmission. With incidence
from the DBR, on the other hand, a finite number of layers enables
optimum coupling to the TP, while the metal may be optically
opaque. These two different regimes for perfect absorption are further explored in the
following section, with comprehensive numerical simulations.

\section{Conditions for complete absorption}
\begin{figure}[!Htbp]
\centering
\includegraphics[width=\columnwidth]{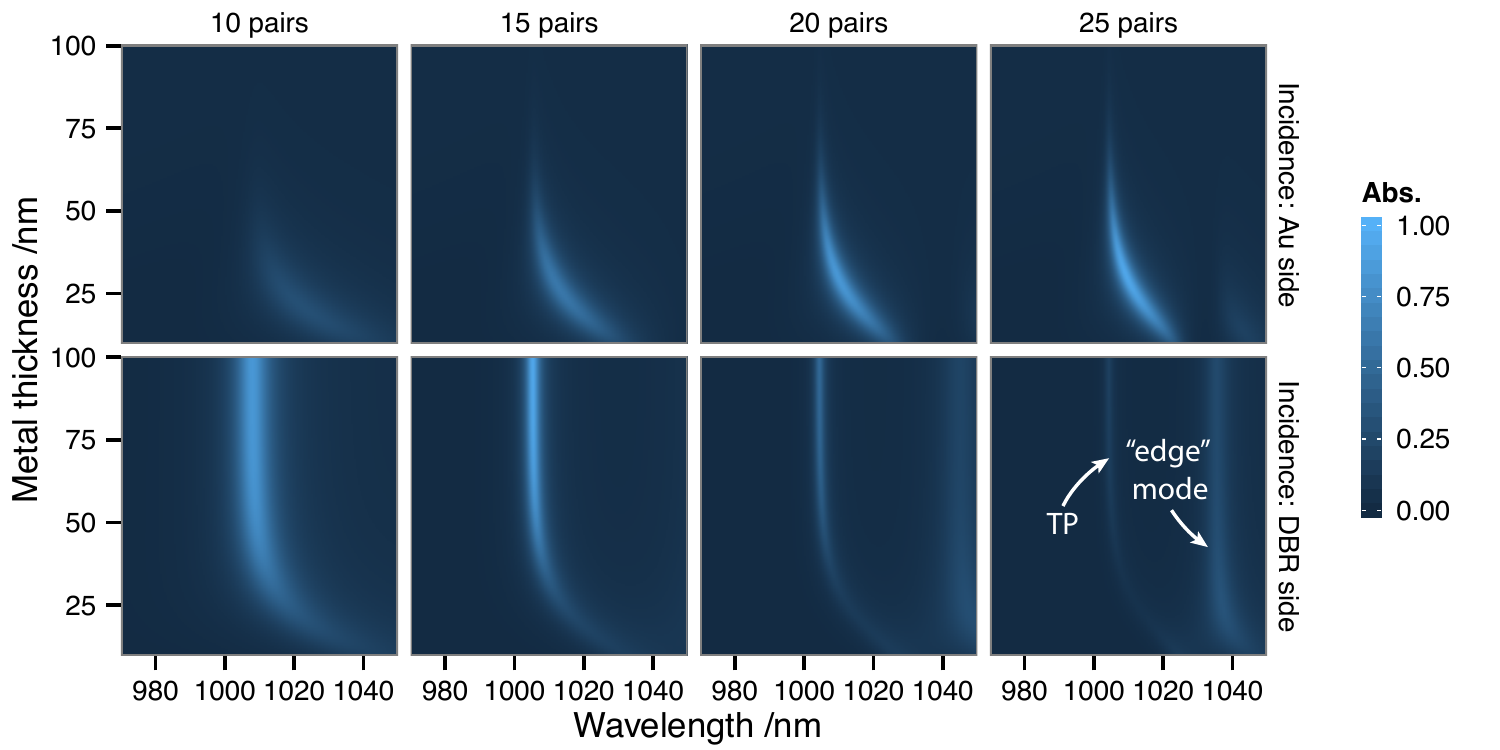}
\caption{Evolution of the absorbance in the region of Tamm plasmon
  excitation when the metal thickness is varied from 20 to 100nm. The
  top and bottom panels correspond to opposite direction of incident
  light (top: air--Au side, bottom: substrate-DBR). The number of periods in the DBR is varied in the panel columns, from 10 to 25 pairs of layers.}
\label{fig2}
\end{figure}

Whether light is incident from one side or the other of the Tamm
structure is irrelevant for the transmission of light, following the
general principle of optical reciprocity. However, the balance
between absorption and reflection can be substantially affected~\cite{Du12}. The
map of absorption shown in Fig.~\ref{fig2} illustrates this asymmetry with respect to
the optimal number of DBR layers, and metal thickness.

Complete absorption signals a large electric field inside the metal
layer, because all other layers have a purely real dielectric function.
Since the Tamm mode is confined at the interface between the metal and
DBR, the condition for maximum absorption coincides with a larger
field enhancement in the nanostructure, 
with the proviso that the spatial field distribution be unaltered. Optimisation of the field enhancement is crucial to many
potential applications~\cite{Mai07}, such as surface-enhanced spectroscopies~\cite{Ru09,Bad14}, refractive index
sensing~\cite{Hom99, Aug14}, non-linear optics~\cite{Afi13,Xue13,Lee13}, opto-mechanical
coupling~\cite{Lan07,Fai13}. From the map shown in Fig.~\ref{fig2}, we note two different conditions for complete absorption.
\subsection{Incidence from the metal side}

In this configuration, light needs to get through the metal layer to
enter the structure. The field penetration into the metal layer dies off
very quickly, with a typical skin depth of about 20\,nm, requiring a thin metal layer for optimal coupling (Fig.
\ref{fig2}). The DBR should be infinite (\textgreater{}30 pairs
suffice, in practice, for these two high-contrast materials), while
  the metal layer has an optimum thickness of about 30\,nm. 
\subsection{Incidence from the DBR side}
In this case, light enters the structure from the DBR side, the metal
thickness being infinite (for all practical purposes,
\textgreater{}200\,nm of Au is opaque to IR light). The DBR should however have a specific number
of layers, 13 pairs for this configuration, such that the in- and
out-coupling of light to the Tamm mode balances precisely the Joule loss in
the metal, yielding equality of radiative and non-radiative dissipation rates for
the Tamm mode. Radiation leakage, responsible for
$\Gamma_\text{rad}$, is directly linked to the residual transmissivity
of the DBR. \\

\begin{figure*}[htbp]
\centering
\includegraphics[width=\columnwidth]{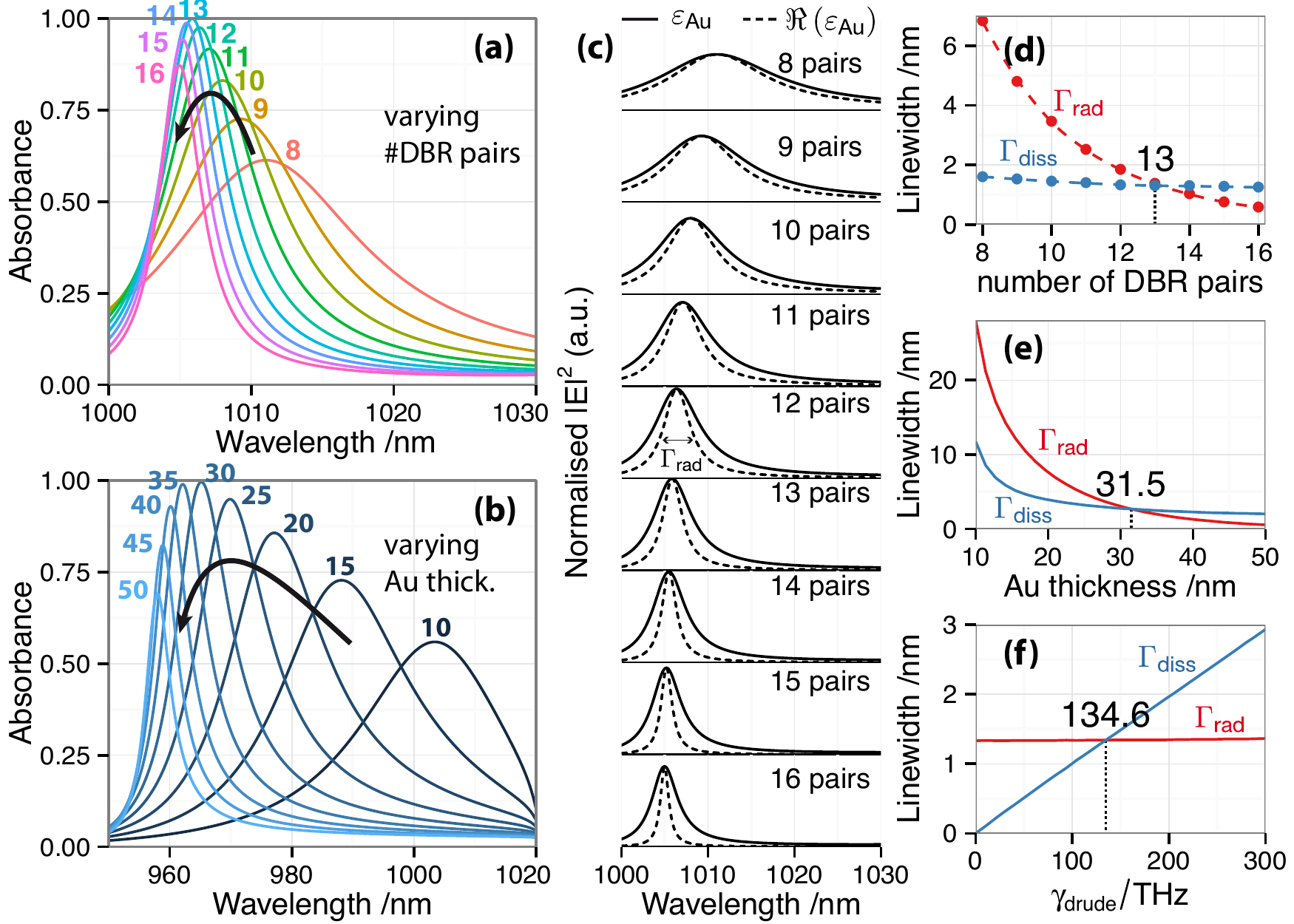}
\caption{Optimisation of the absorbance for the Tamm structure. (a)
  Illumination from the substrate, with opaque Au film. Varying the
  number of DBR pairs from 8 to 16, peak absorbance reaches unity for 13 pairs. (b) With incidence from the air side, and a large
  number of DBR pairs (50), the absorbance is optimised for a metal
  thickness of about 30\,nm. Note that the spacing layer was slightly
adjusted ($s=80$\,nm) in order to move the TP away from the
long-wavelength edge of the DBR stopband. (c) Internal field spectra for
the structure in (a), presenting the electric field intensity at a
fixed position (the middle of the dielectric spacer adjacent to
the metal). Two simulations are presented: the metal film is made of gold
(solid line), or a fictitious material for which the imaginary part of
the dielectric function of gold is set to zero (dashed lines). The intensity is normalised
between the two cases to facilitate the comparison of
linewidths. (d--f) Evolution of the radiative ($\Gamma_\text{rad}$, red) and non-radiative 
($\Gamma_\text{diss}$, blue) contributions to the total linewidth as a function of number of DBR
pairs (d), metal thickness (e), and Drude scattering rate (f).}
\label{fig3}
\end{figure*}

To illustrate the predictive power and physical insight provided by the critical coupling
argument, we isolated the radiative and non-radiative
damping contributions to the TP linewidth using numerical
simulations. The procedure outlined below is only strictly valid for well-defined
resonances (high quality factor), as is the case for the range of parameters considered in this work. Numerical simulations were performed to calculate
the far-field spectra (R, T, A) of various Tamm structures, close to
the perfect absorption condition. The internal electric
field intensity at a fixed position, chosen as the middle of the
spacing layer, follows the same spectral profile as the far-field
properties. If the gold layer is replaced by an artificial material
with the imaginary part of its dielectric function set to zero,
absorption cannot occur anywhere in the structure. Somewhat
counter-intuitively, energy conservation dictates that the
reflectance be unity across the stopband and in particular at the TP resonance (no absorption, and
transmission is almost entirely blocked by the DBR), and only the internal electric field
intensity signals coupling to the localised mode. The linewidth of
this intensity spectrum has for only contribution the radiative loss
suffered by the Tamm mode. The procedure and its results are
shown in Fig.~\ref{fig3} for different configurations of
complete absorption. In panel (a), we vary the number of DBR pairs
for a Tamm structure with opaque gold film, illuminated from the
substrate. As we discussed above, the absorbance reaches unity for an
optimum value of 13 pairs of layers. Panel (b) considers the opposite
situation, where a semi-infinite DBR is terminated by a thin Au film
(light incident from the air side). The spacer thickness was adjusted
to shift the TP resonance closer to the centre of the stopband,
thereby limiting the lineshape distortion above 1020\,nm. Again, we
observe the existence of an optimum parameter, the metal thickness, of
about 30\,nm. Panel (c) illustrates the procedure used to decompose the
radiative and non-radiative contributions to the spectral linewidth,
for the same set of parameters as in Fig.~\ref{fig3} (a). Two sets of
simulations are presented, where the dielectric function of gold is
kept as-is (solid lines), or stripped from its imaginary component
(dashed lines) to remove the contribution of non-radiative damping (Joule
heating,
$\Gamma_\text{diss}\propto\int_\text{volume}\Im(\varepsilon)|E|^2dV$,
with $\Im(\varepsilon)$ the imaginary part of the dielectric function). The
second set of spectra (labeled $\Re (\varepsilon_{Au})$) was normalised to the maximum intensity for the
corresponding (complex) gold simulation ($\varepsilon_{Au}$), to facilitate visual comparison of
the linewidths. A maximum intensity is obtained for 13 pairs of
layers, coinciding with the peak absorbance in panel (a), and the
critical coupling condition $\Gamma_\text{rad} = \Gamma_\text{diss}$. This
equality is further detailed in panel (d), where the two contributions
$\Gamma_\text{rad}$ and $\Gamma_\text{diss}$ (estimated from $\Gamma_\text{tot}-\Gamma_\text{rad}$) are plotted against the number of
DBR pairs. Starting with 8 pairs, the resonance width is dominated by
radiative damping: the DBR is too small a barrier and light
over couples to the Tamm mode. Increasing the number of pairs beyond
the optimum condition sees the linewidth dominated by a non-radiative
component; the Tamm mode is well-confined and mostly subject to ohmic
dissipation. This intrinsic loss characterises the portion of
electromagnetic field localised in the metal, and varies relatively
little in this range of parameters as the modal field distribution is
not strongly affected by the number of pairs in the DBR. 
In the limit of infinite DBR, damping would be
purely non-radiative, but incident light would be unable to couple
to the Tamm mode. Similarly, the optimum condition observed in panel (b)
matches the prediction of $\Gamma_\text{rad} = \Gamma_\text{diss}$
highlighted in panel (e) with the optimisation of metal
thickness. The interplay between radiative and non-radiative damping
contributions is more subtle, as the non-radiative component is also noticeably affected by the thickness of the thin Au
film, through which light must enter the structure to couple to the
Tamm mode. Finally, we consider in Fig.~\ref{fig3} (f) the effect of
the metal dielectric function itself, replacing the
experimentally-obtained values for gold with a Drude model, and
varying the electron-scattering rate around its measured value. As
expected, the non-radiative damping contribution scales linearly with
the Drude loss parameter, while the radiative contribution remains
essentially constant. The condition of critical coupling reaches a consistent prediction,
where $\gamma_\text{drude}$ coincides with the best-fit of the Johnson
  and Christy data for $\varepsilon_\text{Au}$ (further
  details on the Drude model and its fit to the data are presented in
  the S.I.). Naturally, in the limit $\gamma_\text{drude}\to 0$,
  damping is purely radiative.
\section{Tunability under critical coupling}
Figure~\ref{fig0} illustrated that the spectral position of the Tamm mode may be tuned by varying the
thickness of the dielectric spacer immediately adjacent to the metal. In
this concluding section, we discuss the conditions under which such tuning may preserve the critical
coupling condition.

\begin{figure}[htbp]
\centering
\includegraphics[width=\columnwidth]{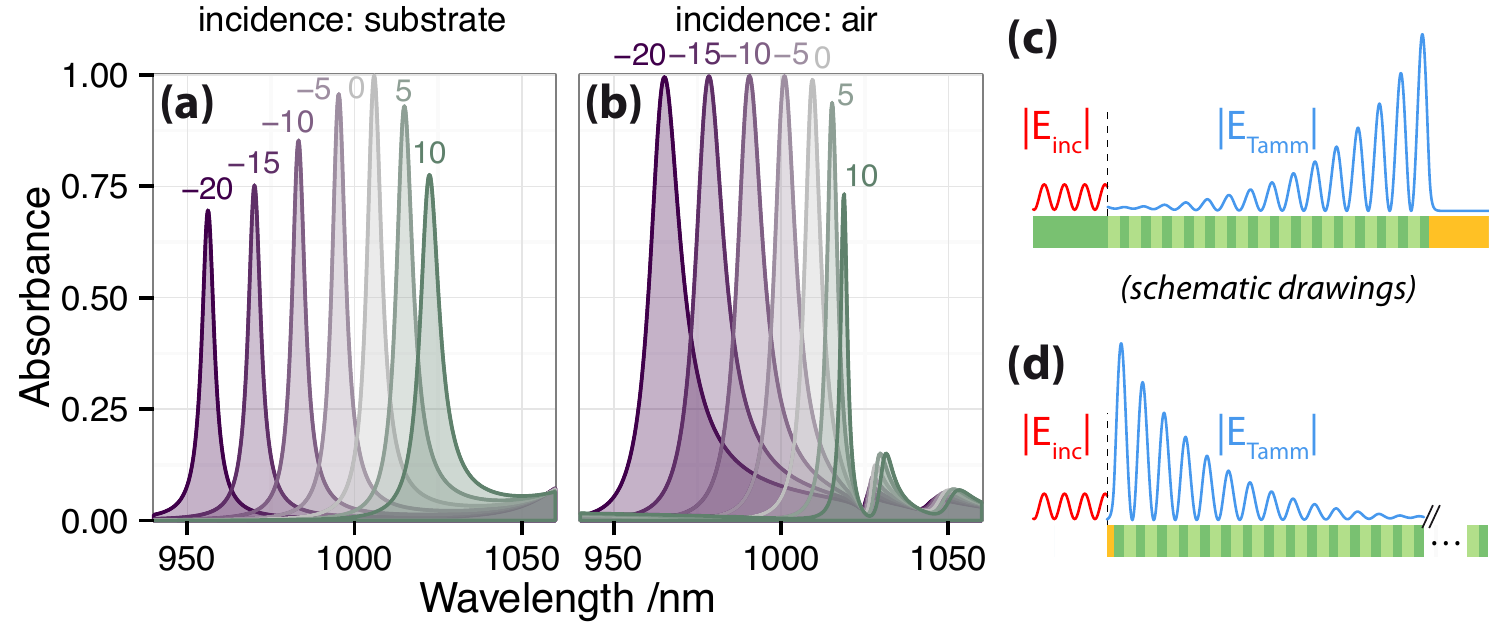}
\caption{Tunability of the TP resonance and its effect on critical
  coupling. (a) Absorbance spectra for a Tamm structure with finite
  DBR and opaque gold film, with light incident from the substrate
  side. The spacer layer thickness is varied around its nominal
  thickness $\lambda/4\approx 64$\,nm from in the range
  $[-20;+10]$\,nm. Only the nominal thickness presents complete
  absorbance. (b) Absorbance spectra corresponding to a practically semi-infinite
  DBR structure (50 pairs) terminated by a 30\,nm-thin Au film, with
  incidence from the air side. The spacer thickness variation shifts
  the Tamm resonance across the stopband, while maintaining 
  unit absorbance. (c) and (d) Schematic drawings of the spatial
  distribution of the Tamm modal field relative to the incident field
  (shown with arbitrary phase), in both configurations.}
\label{fig4}
\end{figure}

A marked contrast is observed in Fig.~\ref{fig4} for the two
configurations: light incident from the substrate with a finite DBR,
(a); light incident from the air side with a thin Au film, (b). In
Fig.~\ref{fig4} (a) the thickness of the spacer has a strong
detrimental effect on critical coupling. How can we understand this
difference with the opposite configuration shown in Fig.~\ref{fig0}(c)
and repeated in Fig.~\ref{fig4} (b), where tuning of the TP over a
wide spectral range can be obtained while maintaining the critical
coupling (and perfect absorbance condition)? We use an analogy with
the process of critical coupling to surface plasmon-polaritons (SPPs)
described in Ref.~\cite{Her94} to interpret this behaviour. In the
Kretschmann (or Otto) configuration, light couples to propagating SPPs
via attenuated total internal reflection, and the observed drop in reflectivity can be
interpreted as a quantum interference between two indistinguishable
pathways: i) total internal reflection, ii) conversion of light to
SPPs, followed by re-radiation in the prism (radiative decay). For the
Tamm structure depicted in the schematics of Fig.~\ref{fig4}(c, d) the
two interfering pathways would be i) direct reflection from the first mirror
(DBR for Fig.~\ref{fig4}(c), Au for Fig.~\ref{fig4}(d)); ii)
conversion to TPs, followed by re-radiation (with probability
proportional to $\Gamma_\text{rad}$). Naturally, the condition of critical coupling
requires an exact match of amplitude and phase between the two
pathways. We note at this point that the modal field associated with
Tamm plasmons is remarkably insensitive to a small variation of the
spacer thickness (further simulations are shown in the S.I.). Thus,
when light enters the structure from the air-Au side, a constant phase
relationship is maintained between the incident light and the Tamm
mode. In contrast, with incidence from the substrate-DBR side, the tail of the Tamm mode which couples to free
radiation undergoes an important phase shift as the first DBR
interface is displaced with the spacer thickness (Fig.~\ref{fig4} c). 

In order to maintain a condition of optimal coupling, the DBR
structure in Fig.~\ref{fig4} (a, c) should undergo concomitant changes in the number of
layers together with the spacer variation. In other words, a specific
number of layers only allows a limited range
of tunability of the critically-coupled TP mode for this direction of incidence.

\section{Conclusions}
We have presented and explained conditions for optimised coupling between light
and Tamm plasmons excited at normal incidence from either side of the
Au--DBR structure, using the powerful concept of critical
coupling. With comprehensive numerical simulations, we elucidated the
asymmetry in reflection and absorption with respect to the direction
of incident light, and found parameters that
yield complete absorption in both configurations. The signature of
critical coupling was confirmed in the equality of absorptive and
radiative dissipation rates. These physical insights may be used in
the design of optimised narrow-band tunable absorbers and thermal emitters.
\section*{Acknowledgments}
B.A. wishes to thank Jean-Jacques Greffet and Christophe Sauvan for fruitful
discussions at the Institut d'Optique in Orsay. 
\end{document}